\documentstyle[amssymb,12pt,thmsa,sw20aip]{article}

\input{tcilatex}
\begin{document}

\author{M.B.Altaie\thanks{%
E-mail: maltaie@yu.edu.jo} \and Dept. of Physics, Yarmouk University, 21163
Irbid-Jordan}
\title{Quantum black hole inflation}
\date{May 2001}
\maketitle

\begin{abstract}
In this paper we follow a new approach for particle creation by a localized
strong gravitational field. The approach is based on a definition of the
physical vacuum drawn from Heisenberg uncertainty principle. Using the fact
that the gravitational field red-shifts the frequency modes of the vacuum, a
condition on the minimum strength of the gravitational field required to
achieve real particle creation is derived. Application of this requirement
on a Schwartzchid black hole resulted in deducing an upper limit on the
region, outside the event horizon, where real particles can be created.
Using this regional upper limit, and considering particle creation by black
holes as a consequence of the Casimir effect, with the assumption that the
created quanta are to be added to the initial energy, we deduce a natural
power law for the development of the event horizon, and consequently a
logarithmic law for the area spectrum of an inflating black hole.
Application of the results on a cosmological model shows that if we start
with a Planck-dimensional black hole, then through the process of particle
creation we end up with a universe having the presently estimated critical
density. Such a universe will be in a state of eternal inflation.
\end{abstract}

\section{Introduction }

Since the pioneering works of Parker [1] and of Zeldovich and Starobinsky
[2], large number of papers has been published on the the subject of
particle creation in curved spacetime. In this respect the celebrated paper
of Hawking [3] marked a milestone in an approach relating gravity with
quantum field theory and thermodynamics. It was discovered that black holes
can creat particles and therefore radiate energy which was found to have a
black-body spectrum, this was later called the `Hawking effect`. Davies [4]
showed that a uniformly accelerated mirror in Rindler space creates thermal
spectrum. Also, it was shown by Unruh [5] that even a uniformly accelerated
observer in Minkowski vacuum experience a thermal bath of created particles,
this was later called the `Unruh effect`.

The detailed analysis of particle creation by a localized gravitational
field (or by accelerated mirrors or Minkowski observers) stems, in essence,
from the fact that the gravitational field (or accelerated systems)
red-shifts the outgoing modes that are leaving the collapsing body in the
Hawking effect. In the case of moving mirror, the Doppler shift reproduces
the same effect that the gravitational field produces on the field modes.
This is the standard explanation given to the phenomena of particle creation
by gravitational field, (for more details see ref. [6] and for more recent
review see ref. [7]). However, the direction of the out-going modes is
another matter, which is normally related with the boundary conditions
considered [7]. Also, whether the spectrum of the emitted radiations will be
thermal or not is a matter of system conditions and the special ansatz
employed [8].

Before the Hawking discovery, Price [9] studied the nonspherical
perturbations of relativistic collapse in presence of a scalar field. He
showed that at certain values of $r$, outside the event horizon there exist
potential barriers with respect to the scalar waves. The positions of these
barriers depend on the value of the multipole $l$. for{\it \ }$l=0$ the
position of the potential barrier is at $r=\frac{8}{3}M$, for $l=1$ it is at 
$r=2.88M$, for $l=3$ it is at $r=2.95M$, and as $l\longrightarrow \infty $
the position of the potential barrier $r\longrightarrow 3M.$

Nugayev and Bashkov [10] considered particle creation by black holes as a
consequence of the Casimir effect. They showed that the gravitational field
of the black hole produces a potential barrier outside the event horizon at $%
r=3M$. In a later paper Nugayev [11] showed that the black hole radiation is
due to the interaction of virtual particles with the ''cavity'' formed
between the event horizon and the potential barrier at $r=3M$. This result
established a close relationship between the Casimir effect and particle
creation by black holes.

This relationship was explicitly treated by Anderson etal [12] who studied
the semiclassical corrections of the scalar field to the Schwartzchild
metric. Using a general form for the energy-momentum tensor which was
calculated by the authors in an earlier work [13], the authors found some
interesting proprties for the vacuum expectation value of the components of
the energy-momentum tensor in the vicinity of the event horizon of a
Scwartzchild black hole. For example they found that the energy density of
the scalar field at the horizon will be negative for all values of the
curvature coupling with values $\xi <4/15$. This means that for $\xi $=$%
\frac{1}{6}$ (the conformally coupled field), and for $\xi =0$ (the
minimally coupled field), the energy density is negative. On the other hand,
they find that the energy density become positive for large values of $\xi $%
.The energy density is found to be everywhere positive outside the horizon
for values of $\xi $ in the range $\frac{4}{15}<\xi <1.2515.$ However, for
large value of $\xi $ outside the horizon, Anderson et al. found a point at
which the energy density is independent of $\xi .$

Berezin, Boyarsky and Neronov [14] considered a massive self-gravitating
shell as a model for the collapsing body and a null self-gravitating shell
as a model for the Hawking radiation. They showed that the mass-energy
spectra for the body and the radiation do not match. They resolved this
situation by considering the resultant effects of the outgoing and incomming
radiation modes together. This consideration led them to the important
result that the quanta of radiation created by a black hole come in pairs,
one emitted to infinity and the other falls on the black hole. This will
obviously cause a change of the inner structure of the black hole resulting,
according to their conclusion, in a Bekenstein-Mukhanov [15] spectrum for
large masses. This conclusion will be discussed futher in this paper.

Recently Sch\"{u}tzhold [16] suggested a cannonical particle definition via
the diagonalization of the Hamiltonian for quantum fields in specific curved
spacetimes. Within his suggested approach radial in-going or out-going
Minkowski particles do not exist. An application of this approach to the
Rindler metric recovers the Unruh effect. The application of the same
approach on particle creation by a black hole shows a genuine dependance of
the Hawking temperature on the dynamics of the collapse. Furthermore it was
shown through the evoluation of the vacuum expectation value of the
energy-momentum tensor that there is no late-time Hawking radiation and
therefore no black hole evaporation.

These investigations, and much of the controvercies over the subject brings
up the need for more elaborate and simple approach to deduce particle
creation by gravitational field. Physically the idea is feasible in the
light of basic knowledge gained from quantum mechanics and general
relativity. Since time is dilated by the presence of gravitational field (or
equivalently the energy spectrum is red-shifted) with respect to a
Minkowskian frame of reference the law of conservation of energy (or the
principle of general covariance in more accurate terms) should allow for
particle creation in pairs that should uphold the conservation laws.

Throughout all the original derivations of particle creation by black holes
or moving boundaries we notice that the main role is played implicitly by
the action of the red-shift of the outgoing frequency modes with respect to
the in-comming frequency modes. This is just another facet of time-dilation
experienced by events taking place in non-inertial frames of reference. This
effect is best clarified by the response function where the red-shift is
made promenant [6].

In this paper we consider a new approach to deduce particle creation by a
localized strong gravitational field. The approach is based on a definition
of physical vacuum, (cosequently a definition of real particle) based on
Heisenberg uncertainty principle. The gravitational red-shift of the
frequency modes is used to determine the minimum strength of the
gravitational field needed to achieve the minimum amount of red-shift
required to convert virtual particles into real ones. Accordingly, a
regional upper limit in the vicinity of the Schwartzchild black hole is
deduced, below which only, real particle creation can take place. This upper
limit is further utilized to define a Casimir second surface in a system
where the first surface is assumed to be the event horizon itself. The
system will induce a non-zero Casimir energy which is assumed to be added to
the initial total energy of the black hole. This addition will cause a
change in the internal structure of the black hole, consequently the area of
the event horizon will increase. This mechanism will cause the black hole to
expand logarithmically, i.e, to inflate indefinately without end. An
application of these results on a model black hole that is assumed initially
to have Planck dimensions leads to a black hole having a mass that is equal
to the critical mass in the present universe. Accordingly we speculate that
if our universe was born as a black hole then it is most possible that it
has the critical mass density by now.

The rest of this paper is organized as follows: in Sec. II we present an
overview of the customary definition of vaccum according to quantum field
theory, and suggest an altarnative definition, which may be less umbiguous,
based on the general relativity and the Heisenberg uncertainty principle. In
Sec.III we investigate the relation between the energy and time in presence
of a localized gravitational field, through the study of the gravitational
red-shift, where we deduce a lower limit on the value of the red-shift
necessary for the conversion of virtual particles of the Minkowski vacuum
into real particles. In Sec. IV we apply this lower limit to a Schwartzchild
black hole where we deduce an upper limit on the region in the vicinity of
the black hole where real particles can be created. This upper limit is
found to coincide with the position of the potential barrier of Price [9].
In Sec.V we setup a Casimir system whithin which the creation of vacuum
energy takes place, and in sec.VI a succssesively developing Cauchy surfaces
structure is constructed to explain the development of the black hole if the
created Casimir energy is to be added to the initial energy. A cosmological
application is presented in sec.VII, where we find that the proposed natural
mechanism which sets the black hole into an inflationary state can develop a
Planck-sized black hole into a universe-sized one with the present critical
density. Finally we present in sec. VIII a thorough discussion of the
results in view of available related investigations, outlining at the same
time further implications of our results. Throughout this paper we use the
natural system of units defined by $\hbar =c=G=1$ unless otherwise stated.

\section{Definition of the Vacuum}

In quantum field theory the vacuum is understood to be the state of no
particle. Although the definition of vacuum in Minkowski spacetime seems to
be trivial, the case is not so in curved spacetime [6,17]. In Minkowski flat
spacetime the vacuum state is defined by

\begin{equation}
\hat{a}|0>=0,  \label{q1}
\end{equation}
where $\hat{a}$ is called the annihilation operator.

The definition in (\ref{q1}) above leads to problems when calculating the
vacuum expectation value of the Hamiltonian $\hat{H}$ for a bound system
(e.g harmonic oscillator) where it is known that

\begin{equation}
<0|\hat{H}|0>=\stackunder{k}{\sum }\frac{1}{2}\omega _{k}  \label{q5}
\end{equation}

The sum on the RHS diverges. These divergences are considered a chronic
disorder in the structure of quantum field theory [18]. Usually special
prescriptions are used to remove these divergences but all lack the
conceptual foundations.

An alternative definition of the vacuum may be obtained by incorporating the
concepts of the general theory of relativity and quantum mechanics. It is
known that the Einstein field equation for empty spacetime,

\begin{equation}
R_{\mu \nu }-\frac{1}{2}g_{\mu \nu }R=0,  \label{q6}
\end{equation}
admits solutions other than the Minkowski flat spacetime solution. It is
notable that all non-flat solutions do contribute a non-zero vacuum energy
density , (called Casimir energy) whereas the flat Minkowski spacetime
contributes nothing. This suggest that a true vacuum state cannot exist in a
curved spacetime, a point which was rigorously supported by the findings of
Fulling [17]. Therefore, and in order to investigate the behavior of matter
fields in curved spacetime, the generally adopted trend resorts to assume
the existence of two asymptotically flat regions, one in the remote past
(past infinity) and the other at late times (future infinity).This is the
approach followed by the references cited above. However, from a general
relativistic point of view, the true vacuum can only exist in the form of
the Minkowski spacetime. This accurately means that a true Minkowski
spacetime has to be physically empty. Therefore, in the conceptual sence, no
particle should be assumed to exist in a Minkowski spacetime, and once a
particle is assumed to exist, the spacetime is no longer Minkowskian.

This bring us to the situation where the vacuum needs to be re-defined. For
this goal, we define the vacuum as being the state of no measurable physical
energy. This means that assumption that a state $|0>$ can exist in the
remote past or in the far future can be replaced by the requirement that

\begin{equation}
Et<1  \label{q7}
\end{equation}
where $E$ is the total energy of the state and$\ t\,$is the duration through
which it render itself for physical measurement.

The condition in (\ref{q7}) is nessessary to assure that no real particle
exist, on the other hand this condition allows for virtual states to be
assumed to exist. Such states are well-known in flat space quantum field
theory and are used to explain some properties of forces and interactions.
Therefore instead of talking about the Minkowski vacuum, we may talk about a
''Heisenberg vacuum'' defined by Eq. (\ref{q7}) where we have now a virtual
particle of energy $E$ allowed to exist within a time interval $t$ such that
Eq. (\ref{q7}) is satisfied.

\section{Energy and Time}

According to the theory of general relativity, if a frequency mode ( a
quanta) of frequency $\omega \,_{2}\,$is emitted in a static gravitational
field at a point $x_{2}$ then this mode will be received at a point $x_{1%
\text{ }}$with a frequency $\omega $ $_{1}$. The relation between the two
frequencies is given by

\begin{equation}
\frac{\omega _{1}}{\omega _{2}}=\left[ \frac{g_{00}(x_{1})}{g_{00}(x_{2})}%
\right] ^{1/2}.  \label{q8}
\end{equation}

This correspond to a blue-shift or a red-shift of the respective frequencies
depending on the relative positions of the points $x_{1}$ and $x_{2}$. If $%
x_{1}$is at the weak field point then an observer at $x_{2}$ will see the
frequency coming from $x_{1}$ blue-shifted, but if $x_{1}$ is at the
stronger field then the observer at $x_{2}$ will see the frequency
red-shifted. Let us assume that $x_{1}$ is a point at the asymptotically
flat region of the space, then we can define the red-shift parameter as

\begin{equation}
z=\frac{\Delta t}{t_{1}}=\left( \frac{g_{00}(x_{1})}{g_{00}(x_{2})}\right)
^{1/2}-1,  \label{q9}
\end{equation}
where

\begin{equation}
\Delta t=t_{2}-t_{1}.  \label{q10}
\end{equation}

The energy of a real state at the two points will be related by Eq. (\ref{q8}%
) above.

Now let us look at the two points $x_{1}\,$and $x_{2}$ not as two separate
points in a given static gravitational fields but as the same point
belonging to two different situations of the field. The first, $x_{1}$, is
when the field is off and the other, $x_{2}$, is when the field is suddenly
switched on. This assumption was already used by Sch\"{u}tzhold [16] and
others in the study of particle creation by black holes.

By assuming $x_{1\text{ }}$to be a point at the asymptotically flat region
(i.e Minkowskian region), we are entitled to consider the states in this
region as representing the true physical vacuum. Therefore, in this region
we can assume that

\begin{equation}
E_{1}t_{1}<1,  \label{q11}
\end{equation}
where $E_{1}$and $t_{1}$are the energy and the lifetime of the state
respectively.

Now switching-on the gravitational field will change the status of $x_{1}$%
into $x_{2}$ where real particles may exist with energy $E_{2}$ and a
lifetime $t_{2}$ satisfying the relation

\begin{equation}
\Delta E\Delta t>1,  \label{q12}
\end{equation}
where

\begin{equation}
\Delta E=E_{1}-E_{2},  \label{q13}
\end{equation}
and $\Delta t$ is as in Eq. (\ref{q10}).

Now, with simple algebraic manupilations one can show that

\begin{equation}
\frac{\Delta t}{t_{1}}-\frac{\Delta E}{E_{1}}\gtrsim \frac{1}{E_{1}t_{1}}+%
\frac{E_{2}t_{2}}{E_{1}t_{1}}-1,  \label{q14}
\end{equation}
and by Eq. (\ref{q11}) we have $\frac{1}{E_{1}t_{1}}>1$, also by Eq. (\ref
{q8}) $E_{1}t_{1}=E_{2}t_{2}$, therefore

\begin{equation}
\frac{\Delta t}{t_{1}}>1+\frac{\Delta E}{E_{1}}.  \label{q15}
\end{equation}

Using the definition of the red-shift parameter $z$ as in Eq. (\ref{q9}) we
dedude that for real particle creation to takeplace in a given gravitational
field there should be a lower limit for the value of the gravitational
red-shift parameter given by

\begin{equation}
z>1,  \label{q16}
\end{equation}
and for values of $z${\it \ }$<$ $1$, only virtual particles may exist.

The condition in Eq. (\ref{q16}) can be expresseed in terms of the
gravitational potential between any two points in a gravitational field.
From Eq. (\ref{q9}) and Eq. (\ref{q16}) we get

\begin{equation}
\left[ \frac{g_{00}(x_{!})}{g_{00}(x_{2)}}\right] ^{1/2}-1>1  \label{q17}
\end{equation}

Thus the condition for real particle creation becomes

\begin{equation}
g_{00}(x_{1})>4g_{00}(x_{2})  \label{q18}
\end{equation}

\section{The Schwartzchild black hole}

Now we apply the result of the previous section to a non-rotating
Schwartzchild black hole.

For a spherical star of mass $M$ and radius $R_{b}$ we have

\begin{equation}
g_{00}(x_{2})=1-\frac{2M}{R_{b}}  \label{q19}
\end{equation}

If we consider the position of the observer to be at a point $x_{1}$in an
asymptotically flat region where $g_{00}(x_{1})\approx 1$, then the
inequality in Eq. (\ref{q18}) gives

\begin{equation}
4-\frac{8M}{R_{b}}<1,  \label{q20}
\end{equation}
which means that

\begin{equation}
R_{b}<\frac{8}{3}M.  \label{q21}
\end{equation}

In terms of the Schwartzchild radius of the star $R_{s}=2M$ this condition
can be written as

\begin{equation}
R_{b}<\frac{4}{3}R_{s}.  \label{q22}
\end{equation}

This condition defines a regional limit outside the event horizon, below
which only, real particle creation can take place. This limit coincides with
the position of the potential barrier for the S-wave part of the massless
scalar field in the gravitational field of a non-rotating black hole
calculated by Price [9].

The above regional limit means that the region in the vicinity of a black
hole in which real particle creation can take place is confined by two
Cauchy surfaces; the first is the horizon of the black hole and the second
is a surface located at $R=\frac{4}{3}R_{s}$. In between these two surfaces,
real particles can be created.

\section{Particle Creation versus Casimir effect}

In flat space it was found that the vacuum fluctuations of the
electromagnetic field give rise to an attractive force between two parallel
conducting flat plates, this creates a negative energy density inversely
proportional to the fourth power of the distance between the two plates.
This was discovered by Casimir [19] and was called the Casimir effect.
Application of this effect in closed spacetimes (e.g the Einstein universe)
has shown that it lead to a positive energy density (for example see ref.
[20, 21]). Further consideration of the problem at finite temperatures
resulted in calculating the finite temperature corrections which was shown
to be an important driver for the thermal development of the universe when
considered as a source for the Einstein field equations [22, 23].

Following Nugaev [11], we may exchange the nonrotating black hole with two
spherical conductors one just outside the event horizon but very near to it,
and the other to be consider just below the upper limit $\frac{4}{3}R_{s}\,$%
. These two surfaces will constitute concentric shells, analogous to the
parallel conducting plates. Any amount of energy created through the Casimir
mechanism between the two shells is assumed to be added to the total energy
of the black hole, hence extending the event horizon by an amount that has
to be controled by the conditions defined for the system. This assumption
seems acceptable in the light of the finding of Berezin et al. [14] that
particles are created in pairs of positive energy by the black hole, where
one is emitted to infinity and the other falls on the black hole causing a
change of the inner structure. Then, we immediately deduce that the new
event horizon will have a radius of $\frac{4}{3}R_{s}$, and so the growth of
the event horizon will go on. This means that

\begin{equation}
R_{n}=\left( \frac{4}{3}\right) ^{n}R_{s},  \label{q23}
\end{equation}
where $n=1,2,3....$.

The argument we place for considering discrete eigen values for the radius
of the event horizon is simple and goes as follows: The first Casimir system
which is composed of the event horizon and the barrier surface at $R_{1}=%
\frac{4}{3}R_{s}$ will creat an amount of positive energy (the Casimir
energy) once formed, then if this energy is assumed to be added to the total
energy of the black hole , the event horizon will be extended to a new
position and the second surface of the new Casimir system (the new position
of the barrier) will be at $R_{2}=\frac{4}{3}R_{1}$ and so it goes.

According to the above mechanism, the surface area of the event horizon of
the black holes will graw as

\begin{equation}
A_{n}=\left( \frac{4}{3}\right) ^{2n}A_{s}.  \label{q24}
\end{equation}

This means that the area of the event horizon is quantized. The quantization
law here is much different from the law deduced by Bekenstein and Mukhanov
[17] according to which the spectrum of the surface area of the event
horizon was uniformly spaced. However, using the loop representation of
quantum gravity, Barreira et al. [8] have shown that the Bekenstein-Mukhanov
area quantization spectrum is unrecovarable, consequently they deduce that
the Bekenstein-Mukhanov result is likely to be an artefact of the ansatz
used rather than a robust result.

From (\ref{q23}) it is clear that the expansion of the black hole will be
logarithmic (i.e inflationary). The number of the inflationary foldings is
given by

\begin{equation}
n=8\log \left( \frac{R_{n}}{R_{0}}\right)  \label{q25a}
\end{equation}

\section{A Constant-Time Hypersurface (Cauchy Surface) Structure}

Using the result in (\ref{q23}) above, we can construct a hypothetical
concentric Cauchy surface structure centered at the black hole singularity.
This structure is characterized by the eigenvalues of (\ref{q24}) and a set
of time-like Killing vectors normal to the surfaces. The transition from one
surface to another is associated with translation in time, and consequently
generation of energy. Therefore each surface will represent an energy level
characterizing a black hole of the corresponding mass. The relative temporal
separation between these surfaces is given by the basic relation

\begin{equation}
\frac{t(x_{1})}{t(x_{2})}=\left[ \frac{g_{00}(x_{1})}{g_{00}(x_{2})}\right]
^{-1/2}.  \label{q25}
\end{equation}

It is clear that the temporal separations between surfaces situated near to $%
R_{s}$ are larger than those far away. If $x_{2}$ is a point at the
asymptotically-flat region where $t(x_{2})\equiv t_{\infty },$then we can
write

\begin{eqnarray}
t(x_{n}) &=&t_{\infty }[g_{00}(x_{n})]^{-1/2}  \nonumber \\
&=&t_{\infty }\left[ 1-\left( \frac{3}{4}\right) ^{n}\right] ^{-1/2}.
\label{q26}
\end{eqnarray}

This structure represents an infinite set of hypothetical Cauchy concentric
spherical surfaces surrounding the black hole prior to the natural
development of the black hole's event horizon. However, if the positive
energy created by the black hole within the specified region is to be added
to the black hole total energy, then the mass development will take the
following form

\begin{equation}
M_{n}<\left( \frac{4}{3}\right) ^{n}M_{0},  \label{q27}
\end{equation}
where $M_{0}$ is the inital mass of the black hole.

Energy levels of the above structure have the following separations

\begin{equation}
E_{n+1}-E_{n}<\frac{1}{3}\left( \frac{4}{3}\right) ^{n}E_{0},  \label{q28}
\end{equation}
where $E_{0}$ is the total initial energy of the black hole.

From (\ref{q23}) and (\ref{q27}) we find that the energy (mass) density of
the system will develop according to the inequality

\begin{equation}
\rho _{n}>\left( \frac{3}{4}\right) ^{2n}\rho _{0},  \label{q29}
\end{equation}
where $\rho _{0}$ is the initial density. This means that

\begin{equation}
\log \left( \frac{\rho _{n}}{\rho _{0}}\right) >-0.2498n.  \label{q30}
\end{equation}

This equation determines the development of the density of the non-rotating
black hole under the condition that the created energy is added to its
initial energy.

\section{Cosmological Applications}

If the universe was born as a singularity, then it is unkown how it has
crossed its own event horizon. Although quantum effects may dissipate the
creation singularity, the presently available calculations, which
incorporate quantum fields into the classical curvature, do not indicate the
possibility of a universe born with crossed event horizon. The available
calculations [21,22.23] indicates that the universe was born as a
finite-sized pach with dimensions less than the Schwartzchild radius.This
implies that the universe may have been born as a black hole and is still
is. This idea is not new, and there are a number of investigations that
support it; for example it was already shown long ago by Oppenheimer and
Snyder [24] that the inside of the Schwartzchild solution could be a
Friedmann universe. Moreover it was shown by Pathria [25] that our present
universe may be described as an internal Schwartzchid solution if it has the
critical energy density. More recent investigations [26] based on the
assumption of the existance of a limiting curvature have shown that the
inside of a Schwartzchild black hole can be attached to a de Sitter universe
at some space-like junction which is taken to represent a short transition
layer. Other senarios in which the universe emerges from the interior of a
black hole are also proposed [27-33].

We will now utilize the results of the previous section to construct a model
for the whole universe. The model adopts the results of previous
calculations [21,23] of the back-reaction of the finite temperature
corrections to the vacuum energy density of the photon field in an Einstein
universe. Although the Einstein universe is staic, the conformal relation
with the closed Robertson-Walker universe [34] allow us \ to consider the
results as being of practical interest. In fact, the discrete spectrum
provided here by the inflating black hole model can be cosidered as
representing instantaneous successions of different states of the Einstein
static universe. The results of the back-reaction calculations have shown
that the thermal development of the universe covers two different regimes;
the Casimir regime, which extends over a very small range of the radius but
huge rise of temperature from zero to a maximum of $1.44\times 10^{32}$K at
a radius of $5.5\times 10^{-34}cm$. At this maximum temperature a phase
transition takes place, and the system cross-over to the Planck regime,
where photons get emitted and absorbed freely exhibiting pure black-body
spectrum. At this point one can identify the primodial universe with an
initial energy density $\rho _{i}$ given by

\begin{eqnarray}
\rho _{i} &=&\alpha T^{4}  \nonumber \\
&=&3.\,2528\times 10^{114}erg/cm^{3}.  \label{q31}
\end{eqnarray}

However, the same calculations [23] showed that the Einstein universe
exhibit the presenty measured microwave background temperature of 2.73K at a
radius of $1.83\times 10^{30}cm.$ Consistancy require us to adopt this value
for the radius of the present universe. Therefore, from (\ref{q25a}) we can
write

\begin{equation}
n=508.176.  \label{q32a}
\end{equation}

Accordingly, we deduce from (\ref{q30}) and (\ref{q32a}) that

\begin{equation}
\log \frac{\rho _{n}}{\rho _{i}}\approx -126.9425.  \label{q33}
\end{equation}

Using the value in (\ref{q31}) for $\rho _{i}$ we get

\begin{equation}
\rho _{n}=3.\,7133\times 10^{-13}erg/cm^{3}.  \label{q34}
\end{equation}

\ This result is very close to the radiation density in the present
universe, calculated in reference to the cosmic microwave background.

One may argue that the estimated radius of the present universe (Hubble
length) is $1.38\times 10^{28}cm$ and not $1.83\times 10^{30}cm.$ In such a
case we find that

\begin{equation}
n=508.176.  \label{q35}
\end{equation}

Therefore, from (\ref{q30}) we have

\begin{equation}
\log \frac{\rho _{n}^{*}}{\rho _{i}}\approx -122.773,  \label{q36}
\end{equation}
which means that

\begin{equation}
\rho _{now}^{*}\approx 6.\,0956\times 10^{-30}gm.cm^{-3}.  \label{q37}
\end{equation}
a figure which is very close to the critical matter density which defines a
flat universe.

\section{Discussion and Conclusions}

The present investigation tackeld many aspects of the subject of particle
creation by localized gravitational fields. The aim was to utilize the basic
concepts and principles of the general theory of relativity and quantum
mechanics and present a simple approach to deduce particle creation by
Schwartzchild black holes. The analysis led us to determine an upper limit
for the region, in the vicinity of the black hole, where real quanta can be
produced. On one hand this result comes in support to the school of thought
which suggests that the quanta of the Hawking effect are created in the
vicinity of the black hole [35], and on the other hand this result confine
the creation region into a spherical shell of thickness $\frac{1}{3}R_{s}$
outside the event horizon. We emphasize here that the regional upper limit
is consistent with the results of other authors [9,10,12].

The simple approach followed in this paper led us to a new quantization law
for the area of the event horizon, and consequently into a new area
spectrum. The main features of the new spectrum is that it is logarithmic
and macroscopic, in contrast to the Bekenstein-Mukhanov spectrum [15], which
was linear and microscopic (i.e, Planck dimensional). In fact the
Bekenstein-Mukhanov spectrum cannot be verified observationally because in
practice the spectrum will look continuous for macroscopic black holes.

A logarithic inflation arises naturally in our model as a result of the
Casimir system assumption. Normally such a model will benefit from all the
privileges of inflationary models over the standard big bang model, less
their conceptual problems.This is indeed the case since it was recently
shown by Easson and Brandenberger [27] that a universe born from the
interior of a black hole will not posses many of the problems of the
standard big bang model. In particular the horizon problem, the flatness
problem and the problem of formation of structures are solved naturally.
This may well be the case for a universe formed of the interior of an
inflating black hole. Perhaps this is the most important result that need to
be analyzed further to see if one can draw some observational consequences.

Our assumptions in this paper are strongly supported by the results we
obtained for matter and radiation densities in the present universe. One can
see clearly that starting with a Planck-dimensional black hole universe, the
mechanism of the quantum development of such a hypothetical universe leads
to a universe having the present critical matter density, a point which is
supported by the recent observational investigations of the Boomerang
project [36]. Further analysis and development of this approach by
investigating a Kerr or Reissner-N\"{o}rdstrom black holes will be
interesting, where one may expect the emergence of a different scheme for
area quantization.

\section{References}

\ \ [1] L. Parker, Phys. Rev. {\bf 183}, 1057 (1969).

\ \ [2] Ya. B. Zeldovich and A. A. Starobinskii, JETP {\bf 34}, 1154 (1972).

\ \ [3] S. W. Hawking, Comm. Math. Phys., {\bf 43}, 199 (1975).

\ \ [4] P. C. W. Davies, J. Phys. A: Math. Gen. 8, 609 (1975).

\ \ [5] W. G Unruh, Phys. Rev. D {\bf 14}, 870 (1976).

\ \ [6] N. D. Birrel and P. C. W. Davies, {\it Quantum Fields in Curved Space%
},(Cambridge University Press, Cambridge, England 1982).

\ \ [7] L. H. Ford, ''Quantum Field Theory in Curved Spacetime'',
gr-qc/9707062.

\ \ [8] M. Barreira, M. Carfora and C. Rovelli, Gen. Rel. Grav. {\bf 28},
1293 (1996).

\ \ [9] R. Price, Phys Rev. D {\bf 5}, 2419 (1972).

[10] R. M. Nugayev and V. I. Bashkov, Phys. Lett. A {\bf 69}, 385 (1979).

[11] R. M. Nugayev, Nuovo Cim. B {\bf 86}, 90 (1985).

[12] P. R. Anderson etal, Phys. Rev. D {\bf 50}, 6427 (1994).

[13] P. R. Anderson, W. A. Hiscock, and D. A. Samuel, Phys.Rev.Lett. {\bf 70}%
, 1793,( 1993).

[14] V. A. Berezin, A. M. Boyarsky and A. Yu. Neronov, Phys. Lett. B {\bf 455%
}, 109 (1999).

[15] J. D. Bekenstein and M. A. Mukhanov, Phys. Lett. B {\bf 360}, 7 (1995).

[16] R. Sch\"{u}tzhold, Phys. Rev. D {\bf 63}, 24014 (2001).

[17] S. Fulling, Phys. Rev.\ D {\bf 7}, 2850 (1973).

[18] J. D. Bjorken and S. D. Drell, ''{\it Relativistic Quantum Fields}'' ,
Mac Graw-Hill, New York (1965).

[19] H.G.B. Casimir, Proc. Kon. Ned. Akad. Wet., {\bf 51}, 793, (1948).

[20] L. H. Ford, Phys. Rev. D , , (1975).

[21] M. B. Altaie and J. S. Dowker, Phys. Rev. D {\bf 18,} 3357 (1978).

[22] B. L. Hu, Phys. Lett. B , , (1981).

[23] M. B. Altaie, ''Back-reaction of quantum fields in an Einstein
Universe'', gr-qc/0104100.

[24] J. R. Oppenheimer and H. Snyder, Phys. Rev. {\bf 56, }455 (1939).

[25] R. K. Pathria, Nature {\bf 240}, 5379 (1972).

[26] V. P. Frolov, M. A. Markov and V. F. Mukhanov, Phys. Lett. B {\bf 216},
272 (1989).

[27] D. A. Easson and R. H. Brandenberger, ''Universe Generation from Black
Hole Interior'', hep-th/0103019.

[28] V. Mukhanov and R.Brandenberger, Phys. Rev. Lett. {\bf 68}, 1969 (1992).

[29] R. Brandenberger, V. Mukhanov and A. Sornborger, Phys. Rev. D {\bf 48},
1629 (1993).

[30] M. Trodden, V. Mukhanov and R. Brandenberger, Phys. Lett. B {\bf 316},
483 \thinspace \qquad (1993).

[31] A Tseyltlin and C. Vafa, Nucl. Phys. B {\bf 372}, 443 (1992).

[32] H. Genreith, ''A Black Hole Emerged Universe'', astro-ph/9905317.

[33] R. Daghigh, J. Kapusta and Y. Hosotani, ''False Vaccuum Black Holes and
Universes'', gr-qc/0008006.

[34] G. Kennedy, J. Phys. A11, L77, (1978).

[35] R. M, Nugayev, Phys. Rev. D {\bf 43}, 1195 (1991).

[36] P. de Bernardis et al., '' First results from the Boomerang
experiment'', astro-ph/0011469.

\end{document}